\documentclass[aps,pre, twocolumn]{revtex4-1}
\usepackage{amsfonts}
\usepackage{amsmath}
\usepackage{amssymb}
\usepackage{graphicx}
\usepackage{times}
\usepackage{physics}
\usepackage{dcolumn}
\usepackage{bm}
\usepackage{epsfig}
\usepackage{color}
\usepackage{colordvi}
\usepackage[dvipsnames]{xcolor}
\usepackage{marginnote}
\usepackage{lipsum}
\usepackage{relsize}
\usepackage{accents}
\usepackage{wasysym}  

\usepackage[colorlinks=true,linkcolor=blue,citecolor=red,plainpages=false,pdfpagelabels]
{hyperref}
\usepackage[encapsulated]{CJK}

\usepackage{color}

\newcommand{\bean}{\begin{eqnarray*}}
\newcommand{\eean}{\end{eqnarray*}}
\newcommand{\ba}{\begin{array}}
\newcommand{\ea}{\end{array}}
\newcommand{\be}{\begin{equation}}
\newcommand{\ee}{\end{equation}}
\newcommand{\bea}{\begin{eqnarray}}
\newcommand{\eea}{\end{eqnarray}}

\newcommand{\no}{\nonumber}

\begin{document}

\title{Quantum Harmonic Oscillators with Nonlinear Effective Masses}
\author{Jen-Hsu Chang,$^{1}$ Chun-Yan Lin,$^{2}$  and Ray-Kuang Lee$^{2,3,4}$}
\affiliation{
$^{1}$Graduate School of National Defense, National Defense University, Taoyuan city 335, Taiwan\\
$^{2}$Institute of Photonics Technologies, National Tsing Hua University, Hsinchu 30013, Taiwan\\
$^{3}$Department of Physics, National Tsing Hua University, Hsinchu 30013, Taiwan\\
$^{4}$Physics Division, National Center for Theoretical Sciences, Taipei 10617, Taiwan}
 \email{rklee@ee.nthu.edu.tw}
  

\begin{abstract}
We study the eigen-energy and eigen-function of a quantum particle acquiring the probability density-dependent effective mass (DDEM) in harmonic oscillators.
Instead of discrete eigen-energies, continuous energy spectra are revealed due to the introduction of a nonlinear effective mass.
Analytically, we map this problem into an infinite discrete dynamical system and obtain the  stationary solutions by perturbation theory, along with the proof on the monotonicity in the perturbed eigen-energies.
Numerical results  not only give agreement to the asymptotic solutions stemmed from the expansion of Hermite-Gaussian functions, but  also unveil a family of peakon-like solutions without  linear counterparts.
As nonlinear Schr{\"o}dinger wave equation has served as an important model equation in various sub-fields in physics, our proposed generalized quantum harmonic oscillator opens an unexplored area for quantum particles with nonlinear effective masses. 
\end{abstract}

\maketitle

\section{Introduction}
Quantum harmonic oscillator is the most important model system in quantum mechanics, which remarkably exhibits an exact, analytical solution with discrete (quantized) eigen-energies compared to the predictions of classical counterparts~\cite{QM}.
Instead with a given mass, $m_0$, when particles (electrons or holes) move inside a periodic potential or interact with other identical particles, their motions differ from those in a vacuum, resulting in an {\it effective mass}~\cite{solidstate}. 
With an effective mass, denoted as $m^\ast$, the corresponding Schr{\"o}dinger equation  for a  quantum particle in a one-dimensional harmonic oscillator,  characterized by the spring constant  $k$, has the form:
\begin{eqnarray}
i\hbar\frac{\partial}{\partial t} \Psi(x,t) = \frac{-1}{2m^\ast(x)}\frac{\partial^2}{\partial x^2} \Psi + \frac{k}{2} x^2 \Psi.
\end{eqnarray}
Here $\Psi(x,t)$ is the probability amplitude function projected in the spatial coordinate.
In particular, with a nonuniform composition in potential or particle distributions, a position-dependent effective mass (PDEM)  Schr\"odinger equation has gained much interest for its applications from semiconductors to quantum fluids~\cite{PDEM-1, PDEM-2, PDEM-3, PDEM-4, PDEM-5}. Recently, a PDEM Schr\"odinger  equation exhibiting a similar position-dependence for both the potential and mass was exactly solved~\cite{AIP}. 

With the correspondence between  Schr{\"o}dinger equation  and the paraxial wave equation,  similar concept of position-dependent effects is also studied in the dispersion management optical fiber link~\cite{fiberbook}.
Moreover, in addition to position-dependence, chromatic dispersion may also have  intensity-dependent dispersion (IDD) in the optical domains~\cite{nlindispersion, IDD}. 
IDD, or in general the nonlinear corrections to the chromatic dispersion as a function of the wave intensity, has  arisen in a variety of wave phenomena,  such as shallow water waves~\cite{Whitham, Whitham-book}, acoustic waves in micro-inhomogeneous media~\cite{acoustic},  ultrafast coherent pulses in quantum well waveguide structures~\cite{JMO},  the saturation of atomic-level population~\cite{microwave}, electromagnetically-induced transparency in a chain-$\Lambda$  configuration~\cite{EIT}, or nonlocal nonlinearity mediated by dipole-dipole interactions~\cite{sg}.
Inspired by IDD,  in this work, we consider  a quantum particle acquiring an probability density-dependent effective mass (DDEM), i.e., $m^\ast(\vert\Psi\vert^2)$, in a harmonic potential  described by the following generalized Schr{\"o}dinger equation:
\be    i\hbar \frac{\partial}{\partial t}  \Psi(x, t) = \frac{-1}{2 m_0}(1 +b\vert \Psi \vert^2) \frac{\partial^2}{\partial x^2} \Psi +\frac{k}{2} x^2 \Psi.  \label{ddem}\ee
Here, the DDEM is approximated by assuming $[m^\ast(\vert\Psi\vert^2)]^{-1} \equiv  [m_0(1-b\, \vert\Psi\vert^2)]^{-1} \approx [m_0]^{-1}(1+b\, \vert\Psi\vert^2)$, with the parameter $b$ denoting the contribution from the nonlinear effective mass term.
As one can see, when the nonlinear effective mass term is zero, i.e.,  $b = 0$, Eq.~(\ref{ddem}) is reduced to the well-known scenario for a quantum particle in a parabolic potential.

However, when $b\neq 0 $, instead of the discrete energies, continuous energy spectra are revealed due to the introduction of a nonlinear effective mass. 
Analytical solutions for the corresponding eigen-energy and eigen-function are derived with the help of perturbation theory.
Numerical solutions obtained by directly solving Eq.~(\ref{ddem}) give good agreement to the analytical ones  obtained from the expansion of Hermite-Gaussian  functions.
Moreover, we unveil a family of peakon-like solutions supported by DDEM, which has no counterpart in the linear limit.
Our perturbed solutions and numerical results  for this generalized quantum harmonic oscillator with nonlinear effective masses opens an unexplored area for quantum particles. 

The paper is organized as follows: in Session II, we introduce the quantum harmonic oscillator into this generalized Schr\"odinger equation with nonlinear effective mass and reduce Eq.~(\ref{ddem}) into an infinite dynamical system. Then, by perturbation theory and with the help of the eigen-solutions of  quantum harmonic oscillator, we study the corresponding eigen-energy  with the introduction of DDEM, as a function of the parameter $b$.
The monotonicity of the perturbed eigen-energy is also proved.
In Section III, explicitly, we derive the analytical solutions of eigen-energies and the corresponding wavefunctions for the ground and second-oder excited states in the asymptotical limit
The comparison between analytical solutions and  numerical results is illustrated in Section IV,  demonstrating good agreement on the solutions with a smooth profile,  stemmed from the expansion of Hermite-Gaussian wavefunctions.
A new family of  peakon-like solutions with a discontinuity in  its first-order derivative is also unveiled,  which has no linear counterparts.
Finally, we summarize this work with some perspectives in Conclusion.

\section {Quantum Harmonic Oscillator with DDEM}
Without loss of generality, in the following, we set $\hbar = 1$, $k >  0$, $m_0 = 1$ for the simplicity in tackling Eq.~(\ref{ddem}).
Here, by lookin for the stationary solutions $\Psi(x, t) = \psi(x)\, e^{-i E\, t}$, we consider a family of differential equations parametrized by a continuous DDEM parameter $b \neq 0$ of the form
\be   E\, \psi+ \frac{1}{2} (1 +b\vert \psi \vert^2) \frac{\partial^2}{\partial x^2} \psi - \frac{k}{2} x^2 \psi  = 0,  \label{ha} \ee
where $E$ is the corresponding eigen-energy, $x$ denotes a real variable for the coordinate, and $\psi(x)$ is a square integrable function. This stationary Schr{\"o}dinger wave equation can be seen as a generalized quantum harmonic oscillator.

When $b=0$ and $k =1$, Eq.~(\ref{ha}) becomes the well-known equation for the quantum harmonic oscillator, which supports  eigen-function of the $n$-th order excited state in the  position representation reads~\cite{aw}:
\begin{equation}
\phi_n (x)=\mu_n e^{-x^2/2} H_n(x),
\label{hermity}
\end{equation} 
where  $\mu_n= (2^n n ! \sqrt{\pi} )^{-1/2}$ and $H_n(x)$ is the $n$-th order Hermite polynomial. 
The  corresponding eigen-values are equal to $E_n=n+\frac{1}{2}\,$, for any  $n\in{\mathbb{N}}$. We are interested in finding pairs $(\psi_n,\,E_n)_b\,$ fulfilling Eq.~(\ref{ha}) for  a set $b\neq 0$.

\subsection{Perturbation Theory for Eigen-Energies and Eigen-Functions}
To investigate Eq.~(\ref{ddem}) with $b\neq 0$ (but keeping $k \neq 1$ first), we apply the perturbation theory based on the expansion of the solution on the eigen-function $\phi_n (x)$. That is, 
\be  \Psi(x, t; E)= \sum_{n=0}^{\infty} B_n (t) \phi_n (x).   \label{exp}
\ee
By plugging this expansion into Eq.~(\ref{ddem}), one has 
\begin{widetext}
\bea
&& i \sum_{n=0}^{\infty} \frac{d B_n (t)}{d t} \phi_n (x)+ \frac{1}{2}\sum_{n=0}^{\infty}  B_n (t) \frac{ d^2 \phi_n (x)}{ d x^2}  
 +  \frac{1}{2} \sum_{p=0, q=0}^{\infty} B_p (t) {\bar B_q (t)} \phi_n (x) \phi_j (x)  [b \sum_{n=0}^{\infty}  B_n (t) \frac{ d^2 \phi_n (x)}{ d x^2}]  - \frac{k}{2}x^2 \sum_{n=0}^{\infty} B_n (t) \phi_n (x),  \no \\
&&= i \sum_{n=0}^{\infty} \frac{d B_n (t)}{d t} \phi_n (x)+ \sum_{n=0}^{\infty}  B_n (t) [\frac{1}{2}x^2 \phi_n(x)-E_n \phi_n (x)]   \\
&&\quad+ \sum_{p=0, q=0}^{\infty} B_p (t) {\bar B_q (t)} \phi_n (x) \phi_q (x)  [b \sum_{n=0}^{\infty}  B_n (t) (\frac{1}{2}x^2 \phi_n(x)-E_n \phi_n (x))  ]  - \frac{k}{2} x^2 \sum_{n=0}^{\infty} B_n (t) \phi_n (x),  \no \\
&&= i \sum_{n=0}^{\infty} \frac{d B_n (t)}{d t} \phi_n (x)+(\frac{1}{2}-\frac{k}{2}) \sum_{n=0}^{\infty}  x^2 B_n (t) \phi_n(x)- \sum_{n=0}^{\infty} E_n B_n (t) \phi_n(x) \label{mu} \\
&&\quad- b  \sum_{n=0, p=0, q=0}^{\infty} E_n B_n (t) B_p (t) {\bar B_q(t)} \phi_n (x) \phi_p (x) \phi_q (x) + \frac{b}{2} x^2 \sum_{n=0, p=0, q=0}^{\infty}B_n (t) B_p (t) {\bar B_q (t)} \phi_n (x) \phi_p (x) \phi_q (x)  =0.\no   
\eea 
\end{widetext}
Here, $\bar B_n$ means the complex conjugate of $B_n$.
Then, by multiplying Eq.~(\ref{mu}) with $\phi_m(x)$ and using the orthonormal property of $\phi_m(x)$, we obtain 
\bea
&&b  \sum_{n=0, p=0, q=0}^{\infty} [-E_n V_{m,n,p,q}+\frac{1}{2}W_{n,m,p,q}] B_n (t) B_p (t) {\bar B_q (t)} \no\\
&&+  i  \frac{d B_m (t)}{d t} -  E_m B_m(t)+(\frac{1}{2}-\frac{k}{2}) \sum_{n=0}^{\infty} \Gamma_{m,n} B_n (t) =0,
\label{inf}
\eea
where $ \Gamma_{m,n}$, $ V_{m,n, p,q}$, and $W_{m,n,p,q}$ are defined as:
\bean 
&& \Gamma_{m,n} = \int_{-\infty}^{\infty} \phi_m (x) \phi_n (x)dx,  \\
&&V_{m,n, p,q} =\int_{-\infty}^{\infty} \phi_m (x) \phi_n (x)\phi_p (x) \phi_q (x)dx , \\
&& W_{m,n,p,q} =\int_{-\infty}^{\infty} x^2\phi_m (x) \phi_n (x)\phi_p (x) \phi_q (x)dx. \eean
As one can see from Eq.~(\ref{inf}), now, we reduce the original partial differential equation into the infinite discrete dynamical system~\cite{pk}.
With the help of the recursive relation of Hermite polynomial $H_m(x)$, for example see Ref.~\cite{aw}, i.e., $ x^2 H_m(x)= m(m-1)H_{m-2} (x)+(m+1/2) H_m(x) +1/4 H_{m+2} (x)$, one can arrive at 
\bea 
W_{m,n,p,q} &=&\frac{\sqrt{m(m-1)}}{2}V_{m-2,n, p,q} +(m+1/2) V_{m,n,p,q}\no\\
&+& \frac{\sqrt{(m+1)(m+2)}}{2}V_{m+2,n,p,q}. \label{re} \eea
Along with the fact that the non-zero terms in $\Gamma_{m, n}$ are $\Gamma_{n-2, n}=\frac{\sqrt{n(n-1)}}{2}$, $\Gamma_{n, n}=\frac{2n+1}{2}$, $\Gamma_{n+2, n}=\frac{\sqrt{(n+1)(n+2)}}{2}$, we look for the  stationary solution in the form :
\begin{eqnarray}
\Psi(x, t;E)= e^{-iE\, t}\sum_{n=0}^{\infty} B_n \phi_n (x),
\end{eqnarray}
with $B_n \in  \mathcal{R}$. 
Then, for a given energy value $E$, one yields 
\bea
 &&b  \sum_{n=0, p=0, q=0}^{\infty} (-E_n V_{m,n,p,q}+\frac{1}{2}W_{n,m,p,q}) B_n  B_p  B_q \no \\
 &&+ (E -E_m) B_m+(\frac{1}{2}-\frac{k}{2}) \sum_{n=0}^{\infty} \Gamma_{m,n} B_n  =0.  \label{mu3}  
\eea
From now on, for simplicity, we assume $k=1$. 
With the help of Eq.~(\ref{mu3}), next, we consider the perturbation on the energy deviated from the eigen-energy $E_n$ with the corresponding Hermite-Gaussian eigen-mode $\phi_{n}(x)$.

Similar to the methodology used in dealing with the nonlinear mean field in the Gross-Pitaevskii  equation (GPE)~\cite{am,ap}, we  substitute  $\psi(x;E)= \sqrt{P(E)}\, \phi(x)$  with  $ \left\|  \phi(x)  \right\|=1$ into Eq.~(\ref{ha}) and arrive at a nonlinear eigen-energy equation:
\be 
E\, \phi (x)+\frac{1}{2} [1+b\, P (E)\, \phi^2 (x)]\phi_{xx} (x)- \frac{1}{2} x^2 \phi=0. \label{tub} \ee
Once again,  in Eq.~(\ref{tub}), we can see that if $P(E) \to 0$, then the resulting eigen-energy $E \to E_{n}=n+\frac{1}{2}$.
By substituting  $\phi (x)$, obtained from Eq.~(\ref{mu3}),  into Eq.~(\ref{tub}), one can  have the relation between $E$ and $P(E)$ near  $E_{n}=n+\frac{1}{2}$.
In general, the perturbation approach illustrated above works for all the values of $n$. However, only when $n$ is even,  a neat formula can be conducted by taking the advantage of symmetric wavefunctions in $\psi(x)$. 
For even numbers,  $2n$, the resulting eigen-energy $E$ due to the introduction of the DDEM parameter $b$ can be approximated as
\bea 
E & \approx &  E_{2n} - b\, P(E) [ \frac{1}{2} W_{2n, 2n, 2n, 2n} -(2n+\frac{1}{2}) V_{2n, 2n, 2n, 2n}],   \no\\
&=&  E_{2n} - b\, P(E)\, \mu_{2n}^4\, [ \frac{1}{2} \int_{-\infty}^{\infty} x^2 e^{-x^2} H_{2n}(x)^4 dx \label{ei} \no\\
&&\qquad\quad -(2n+\frac{1}{2}) \int_{-\infty}^{\infty} e^{-x^2} H_{2n}(x)^4 dx ].  \eea
To compute the integrals shown in Eq.~(\ref{ei}),  one can utilize the Feldheim identity  for the Hermite polynomials~\cite{aw}:
\be  H_{m}(x) H_{n}(x)=\sum_{\nu=0}^{min (m,n)} H_{m+n-2\nu} (x) 2^{\nu} \nu ! \left(\ba{c}  m \\ \nu   \ea \right) \left(\ba{c}  n \\ \nu   \ea \right),  \label{pr} \ee
and the Titchmarsh's integral formula~\cite{gr}:
\bea && \int_{-\infty}^{\infty} e^{-2x^2} H_{2m} (x) H_{2n} (x)H_{2p} (x) dx   \label{po} \no \\
&& =\pi^{-1} 2^{m+n+p-1/2}\Gamma(n+p+\frac{1}{2}-m) \times\no\\
&&\quad\Gamma(m+p+\frac{1}{2}-n)  \Gamma(m+n+\frac{1}{2}-p), \eea
where $ n+p \geq m, m+p \geq n$ and  $m+n \geq p$; otherwise, the integral is zero.
From Eqs.~(\ref{pr}) and (\ref{po}), a direct calculation  can  yield 
\bea && \int_{-\infty}^{\infty} e^{-2x^2} H_{2m} (x) H_{2n} (x)^3         dx \\ \no
&& = \frac{1}{\pi}  2^{m+3n-1/2} \sum_{\nu=0}^{min (2m,2n)}  \nu !  \left(\ba{c}  2m \\ \nu   \ea \right) \left(\ba{c}  2n \\ \nu   \ea \right)\times\\ \no
&& \quad \Gamma(n+\nu+1/2-m)\Gamma(m+n+1/2-\nu)^2.
\label{com} \eea

\subsection{Monotonicity in the perturbed eigen-energy}
Given $b > 0$ ($b < 0$), to ensure solutions with linear limit $\psi(x; E) \approx \sqrt{P(E)}\, \phi_n(x)$ to exist only if $E \ge E_n = n +\frac{1}{2}$ ($E \le E_n = n +\frac{1}{2}$), we prove  that
the two integrals inside the square brackets in Eq.~(\ref{ei}) is monotonic, i.e.,
\bea
&&  \frac{1}{2}\int_{-\infty}^{\infty} x^2 e^{-x^2} H_{2n}(x)^4 dx -(2n+\frac{1}{2}) \int_{-\infty}^{\infty} e^{-x^2} H_{2n}(x)^4 dx  \no\\
&&< 0;  \label{un} \eea
or equivalently
\be W_{2n, 2n, 2n, 2n} -(4n+1) V_{2n, 2n, 2n, 2n} <0. \label{une} \ee
for $n \geq 0$. 
In Appendix, the proof on the monotonicity for Eq.~(\ref{un}) and Eq.~(\ref{une}) is given in details.

By using the upper and lower solution method developed in the variational calculus~\cite{am}, we can further prove the existence of a positive solution (node-less state)  through  the corresponding Lagrangian for Eq.~(\ref{ddem}),  i.e.,
\bea  &&\mathcal{L}[\psi(x)] =  \int_{-\infty}^{\infty} \left [(-\frac{E}{b} +\frac{k x^2}{2b} ) \ln \vert 1+b\vert \psi\vert^2 \vert + \frac{1}{2}\vert\psi_x\vert^2 \right] dx, \no \\
&& =  \int_{-\infty}^{\infty} \left [(\frac{kx^2}{2b} ) \ln \vert 1+b\vert\psi\vert^2 \vert + \frac{1}{2}\vert\psi_x\vert^2 \right] dx-E\,Q(E),\label{lg1} \\
&& \equiv Z(E)-E\,Q(E), \label{ag} \eea
where $Z(E)\equiv \int_{-\infty}^{\infty} \left [(\frac{kx^2}{2b} ) \ln \vert 1+b\, \vert\psi\vert^2 \vert + \frac{1}{2}\vert\psi_x\vert^2 \right] dx $.
Here, we also introduce the {\it probability factor} $Q$ for this quantum harmonic oscillator with DDEM, by defining
\be 
Q \equiv \frac{1}{b} \int_{-\infty}^{\infty} \ln \vert 1+b   \vert  \psi \vert^2 \vert \, dx. \label{pwf} \ee
As the original generalized Schr{\"o}dinger equation given in  Eq.~(\ref{ddem}) preserves the $\text{U}(1)$ symmetry, i.e., $ \psi \to \exp[i \theta] \psi$,  the conserved density for this model equation can be derived from Noether theorem~\cite{po}.
It is noted that  Eq.~(\ref{pwf})  is only applicable when  $b \neq 0$.
When $b \ll 1$, this probability factor $Q$ can be approximated as
\be
\lim_{b \ll 1}\, Q \approx \int_{-\infty}^{\infty}  \vert  \psi \vert^2  \, dx, 
\ee 
which is reduced to the standard definition of probability for quantum wavefunctions.
For $b = 0$, the corresponding Lagrangian density given in Eq.~ (\ref{lg1}), as well as the conserved density given in Eq.~ (\ref{pwf}), both go to infinity.

These two terms, $Z(E)$ and $Q(E)$, shown in Eq.~(\ref{ag}), correspond to the Lagrangian of our generalized harmonic oscillator and the conserved quantity, respectively.
As the DDEM parameter $b \to 0$, the Lagrangian shown in  Eq.~(\ref{ag}) can be reduced to \[ \int_{-\infty}^{\infty} (-E\, \vert\psi\vert^2+ \frac{k}{2} x^2 \vert \psi\vert^2 +\frac{1}{2}\vert\psi_x\vert^2 ) dx, \]
which is the Lagrangian for the linear equation, i.e.,  $-\frac{1}{2} \psi_{xx}+\frac{k}{2} x^2\psi=E\, \psi$. 
By following the same concept in tackling weak non-linearity~\cite{ka},  the perturbation theory based on the expansion of the Hermite-Gaussian  functions to deal with the DDEM ensures that when $E \to E_{n}$,  one has $Q(E) \to 0$.

\section{Eigen-Energies and Eigen-functions obtained from perturbation}

\subsection{The Ground State}
Now with the analytical formula give in Eq.~(\ref{ei}), we explicitly give the perturbed eigen-energy $E_0^{b}$ and eigen-function $\psi_0^{b} (x)$ for  the ground state in our generalized quantum harmonic oscillator with a given DDEM parameter $b$.
For the ground state, we can assume that  $B_0 \gg B_{2n}, n=1,2,3, \cdots$.
Then, from Eq.~(\ref{mu3}), one has 
\be B_0^2  \approx \frac{E_0^{b}-E_0 }{b (-\frac{1}{2}W_{0,0,0,0}+\frac{1}{2}V_{0,0,0,0}) },   \label{fi}    \ee
and 
\bea B_{2n} &  \approx & \frac{b B_0^3(-\frac{1}{2}W_{2n,0,0,0}+E_{0}V_{2n,0,0,0}) }{(E_0^{b}-E_{2n}) },  \label{fi2}\\\no
 & =  &  B_0 \frac{   (E_0^{b}-E_0 ) (-\frac{1}{2}W_{2n,0,0,0}+E_{0}V_{2n,0,0,0}) }{(E_0^{b}-E_{2n})(-\frac{1}{2}W_{0,0,0,0}+ \frac{1}{2} V_{0,0,0,0}) },     \eea
where $\Gamma_{m,n}$, $ V_{m,n, p,q}$, and $W_{m,n,p,q}$ have the values:
\bean  
&& W_{0,0,0,0} =\frac{1}{4 \sqrt{2\pi}},\\
&& V_{0,0,0,0} =\frac{1}{\sqrt{2\pi}},\\
 && V_{2n,0,0,0} =\frac{(-1)^{n}}{\sqrt{\pi 2^{2n+1} (2n) !} } (2n-1) !!,  \\
 && W_{2n,0,0,0} = \frac{(-1)^{n+1}}{\sqrt{\pi 2^{2n+5} (2n) !} } (2n-1) !! (2n-1),
\eean 
for $n \geq 1$. Therefore, from Eqs.~(\ref{fi}) and (\ref{fi2}), explicitly we have, noting that $E_0=\frac{1}{2}, E_{n}=n+\frac{1}{2}$, 
\bea B_0^2 &  \approx &  \frac{8 \sqrt{2 \pi} }{3b} ( E_0^{b}-\frac{1}{2} ), \\
B_{2n} &  \approx &b B_0^3 \frac{(-1)^{n}  (\frac{1}{4}n +\frac{3}{8})(2n-1)!!  }{\sqrt{\pi 2^{2n+1} (2n) !}(E_0^{b}-2n-\frac{1}{2}) }, \\
 &  \approx & \frac{1}{8}b B_0^3 \frac{(-1)^{n+1} (2n-1)!! }{\sqrt{\pi 2^{2n+1} (2n) !} }, \quad \text{as} \quad n \to \infty. \label{fm}
\eea
With the coefficients above,   the perturbed solution of $\psi_0^b(x)$ can be conducted immediately  as 
\[   \psi_0^{b} (x) \approx  B_0 \phi_0(x)+ B_2 \phi_2(x) 
+ B_4 \phi_4(x)+ \cdots .   \]
We notice that $ \psi_0^{b} (x) \to \phi_0(x) $ as $E_0^{b} \to E_0 = \frac{1}{2}$.
Again, with the orthonormality  of $\phi_{2n} (x) $, in the asymptotical limit, $n \to \infty$, the probability factor $Q$ defined in Eq.~(\ref{pwf}) becomes:
\bea  Q(E_0^{b}) &=&  \frac{1}{b} \int_{\infty}^{\infty} \ln \vert 1+b  \vert  \psi_0^{b}\vert^2 (x)  \vert dx, \label{sm}\\ \no
& \approx& \int_{-\infty}^{\infty} \vert \psi_0^{b}\vert^2 (x) dx  = B_0^2+ B_2^2 + B_4^2+ \cdots. \\ \no
&\approx &  \frac{8 \sqrt{2 \pi}} {3b} (E_0^{b}-\frac{1}{2})[1+\frac{1}{9}(E_0^{b}-\frac{1}{2})^2 \sum_{n=1}^{\infty} \frac{(2n) !} { (n ! 2^{2n})^2  } ], \eea 
It is noted that the identity $ (2n-1)!!=\frac{(2n)! }{2^n n!}$ is applied.
Therefore, we see that $Q(E_0^{b}) \to 0 $  as $E_0^{b} \to E_0=\frac{1}{2}$. 
As one can see from Eq.~(\ref{sm}), the probability factor $Q(E)$ is linearly proportional, in the leading order,  to the eigen-energy $E$, but with the coefficient inversely proportional to the DDEM parameter $b$.

\subsection{The Second Order Excited State}
In addition to the ground state with $n =0$, in general, all the perturbed eigen-energy $E_{2n}^{b}$ and eigen-function $\psi_{2n}^{b} (x)$ can be written explicitly. Here, we illustrate the solutions for the second order excited state, $E_{2}^{b}$ and $\psi_{2}^{b}(x)$,  by assuming  $B_2 \gg B_{2n}, n=0,2,3, \cdots$.
Again, with Eqs.~(\ref{re}) and (\ref{mu3}), one can directly obtain:  
\be B_2^2  \approx \frac{E_{2}^{b}-E_2 }{b [\frac{-\sqrt{2}}{4}V_{0,2,2,2}- (\frac{5}{4}-E_2) V_{2,2,2,2}-\frac{\sqrt{3}}{2} V_{4,2,2,2}] },    \label{fh}    \ee
and 
\be B_{2n}   \approx  \frac{b B_2^3(-\frac{1}{2}W_{2n,2,2,2}+E_{2}V_{2n,2,2,2}) }{(E_{2}^{b}-E_{2n}) },  \label{fh2}  \ee
with  
\be V_{2n,2,2,2}=\frac{(-1)^{n-3} (2n-1) !! (8n^3-60 n^2+94 n-1)} {\sqrt{\pi 2^{2n+10} (2 n)!}}. \label{pro} \ee
As a result, we we have 
\be B_2^2  \approx \frac{256\sqrt{2\pi}}{327b  }  (E_{2}^{b}-E_2 ),  \label{fh3}    \ee
and
\bea B_{2n}^2  & \approx &   \frac{b^2 B_2^6 (2n) ! (16 n^4 -112n^3+32n^2+476n +23)^2}{2^{4n+16}  \pi (n !)^2(E_{2}^{b}-E_{2n})^2 } ,  \no \\ 
& \approx &  b^2 B_2^6 \frac{ (2n) ! n^6}{ 2^{4n+10} \pi  (n !)^2 },  \qquad\qquad \text{as} \quad n \to \infty. \label{fh4}  \eea
Then, the perturbation of $\psi_2^b (x)$ can be constructed by collection the coefficients above, i.e., 
$\psi_2^b (x) \approx   B_2 \phi_2(x) +B_4 \phi_4(x)+ B_6 \phi_6(x)+ \cdots$.
It is noted that here, the expansion starts from $n = 2$ as $B_0 = 0$.
Again, we have  $ \psi_2^b (x) \to \phi_2(x)$ as $E_{2}^{b} \to E_2=\frac{5}{2}$. 
Moreover, thee resulting probability factor $Q(E_{2}^{b})$ in the asymptotical limit, $n \to \infty$ has the form:
\bea  &&Q(E_{2}^{b}) =  \frac{1}{b} \int_{-\infty}^{\infty} \ln \vert 1+b \vert \psi_2^{b}(x) \vert^2  \vert dx, \label{sm2} \\ \no
&&\approx \int_{-\infty}^{\infty} \vert \psi_2^b\vert^2 (x) dx  =  B_2^2 + B_4^2+ B_6^2 + \cdots. \\ \no
&&\approx   \frac{512 \sqrt{2 \pi}} {435b} (E_{2}^{b}-\frac{5}{2})\times \\ \no
&&\quad   [1 +\frac{2 \times  512^2}{ 435^2}(E_{2}^{b}-\frac{5}{2})^2 \sum_{n=1}^{\infty} \frac{(2n) ! n^6} { (n ! 2^{2n+5})^2  } ]. \eea
Here, again, we see that $Q(E_{2}^{b}) \to 0 $  as $E_{2}^{b} \to E_2=\frac{5}{2}$.  

In addition to the ground and second order excited states, for all the even number of $n$,  the perturbed eigen-energy $E_{2n}^{b}$ and eigen-function $\psi_{2n}^{b} (x)$, as well as the corresponding probability factor $Q(E_{2n}^{b})$,  can be derived explicitly, with the help of Eqs.~(\ref{mu3}), (\ref{ei}) and (\ref{pwf}), respectively.
As for the odd number of $n$, Eqs.~(\ref{mu3}) and (\ref{tub}) provide the required conditions to have the  eigen-energy and eigen-function with introduction of the DDEM parameter $b$.

\begin{figure}[t]
\includegraphics[width=8.6cm]{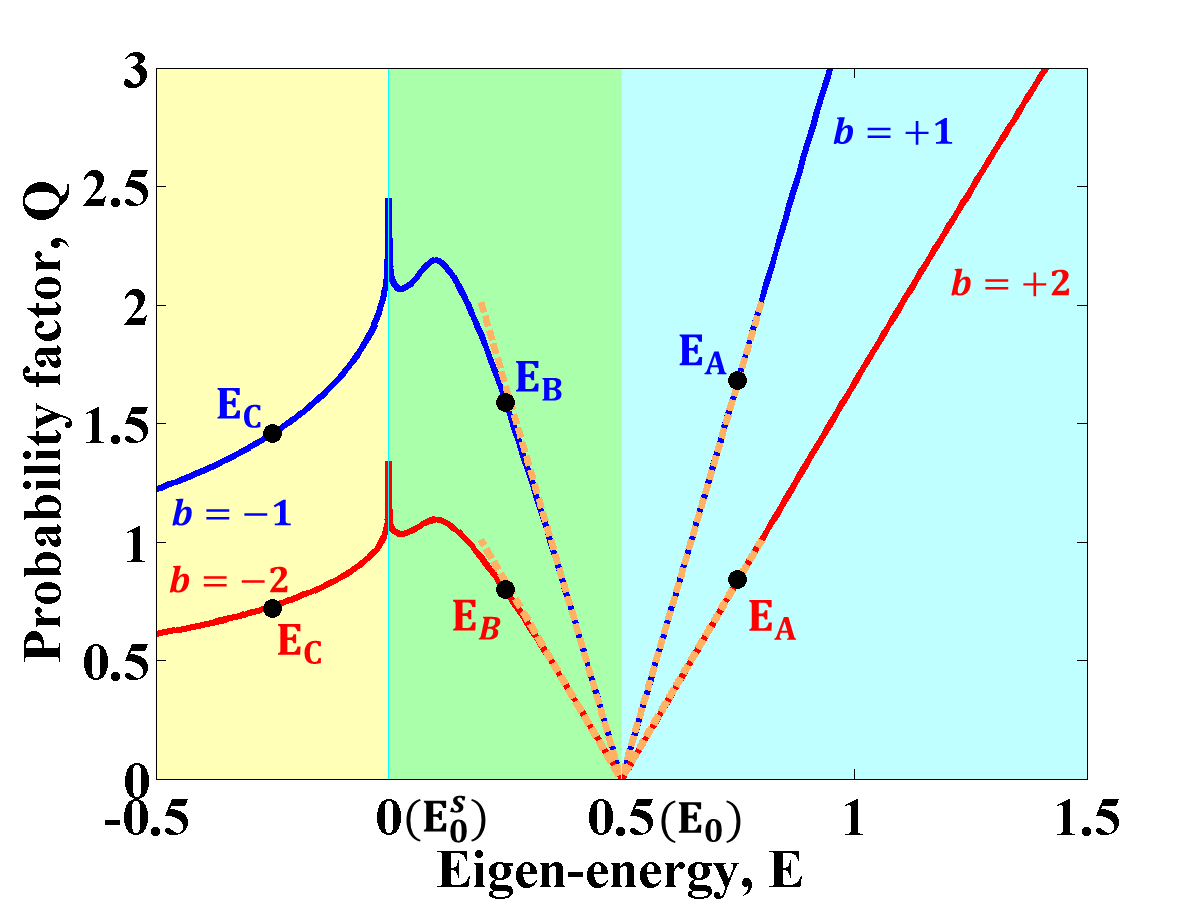}
\caption{The probability factor $Q(E)$ defined in Eq.~(\ref{pwf}) for the ground state in quantum harmonic oscillator with DDEM, as a function of eigen-energy $E$. Here, the DDEM parameter $b$ is set as $\pm 1$ and $\pm 2$,  depicted in Blue- and Red-colors, respectively. The ground state energy in the linear limit, $b = 0$, is marked as $E_0= 0.5$. Analytical solutions based on the perturbation theory given in Eq.~(\ref{sm}) are also depicted in the dashed-curves, which illustrate good agreement to the numerical solutions near $E_0$.
Moreover,  three different regions for the supported eigen-functions are identified for those with a  smooth profile  stemmed from the expansion of  Hermite-Gaussian  wavefunctions (in  Blue- and Green-colored backgrounds for $b > 0$ and $b < 0$, respectively); and with a peakon-like profile having a singularity in  its first-order derivative (in Yellow-color background).}
\end{figure}

\section{Numerical Results by Direct Simulations}

\subsection{The Ground State}
In order to verify the validity of our analytical solutions obtained by the perturbation theory, we also perform the numerical calculations for Eq.~(\ref{ha}) directly without applying any approximation.
To maintain some level of formal rigor and mathematical correctness, we shall talk about finding solutions of differential equations~\cite{PT}.
To find the solutions of the eigen-value problem with the nonlinear term, we  connect with a quantum harmonic oscillator by solving Eq.~(\ref{ha}) with Fourier spectral method. Using the matrix elements, we diagonalize the matrix numerically and perform the iteration to ensure the truncated Fourier basis having the eigen-value converged.
For low energy states, already the smallest basis of $512$ elements gives more than sufficient accuracy.

In Fig. 1, we show the corresponding lowest eigen-mode of the generalized quantum harmonic oscillator described in Eq.~(\ref{ha}), in the plot of probability factor versus eigen-energy $Q$-$E$.
Starting from $E_0 = 0.5$, i.e., the eigen-energy of ground state in the standard quantum harmonic oscillator with $b = 0$, now the eigen-energy is no long a discrete value, but a continuous function due to the introduction of DDEM, i.e., $b\neq 0$.
Here, the initial guess solution has a single-hump profile, i.e., a Gaussian function stemmed from the zero-th order $H_n(x)$.
With a positive value of $b$, such as $b = 1$ and $b =2$, shown in Blue- and Red-colored curves in Fig. 1, the corresponding probability factor $Q(E)$ presents an almost linear function of the eigen-energy $E$.
Now, all the eigen-energy $E_0^b$ are larger than that of $E_0$.
Compared to the analytical formula of $Q(E_0^b)$ obtained in Eq.~(\ref{sm}), the dashed-curves give  agreement to the numerical ones, not only on the slope of $Q$-$E$ curves but also on the inversely proportional dependence on $b$.

Moreover, the corresponding wavefunction $\psi_0^b(x)$ is depicted in Fig. 2(a), which shares a similar Gaussian profile with that in the linear case $b=0$.
For example, at the marked eigen-energy $E_A = 0.75$, the eigen-functions $\psi_0^b(x)$ have similar Gaussian shapes both for $b = 1$ and $b = 2$. 
But with a larger value in the DDEM parameter, such as $b = 2$,  the amplitude, as well as the width, becomes smaller in  the corresponding eigen-functions, as the Red-colored curves shown  in Fig. 2(a).
The analytical solutions obtained by perturbed theory, depicted in dashed-curves in Fig. 2(a), also reflect this similarity. 

\begin{figure}[t]
\includegraphics[width=8.6cm]{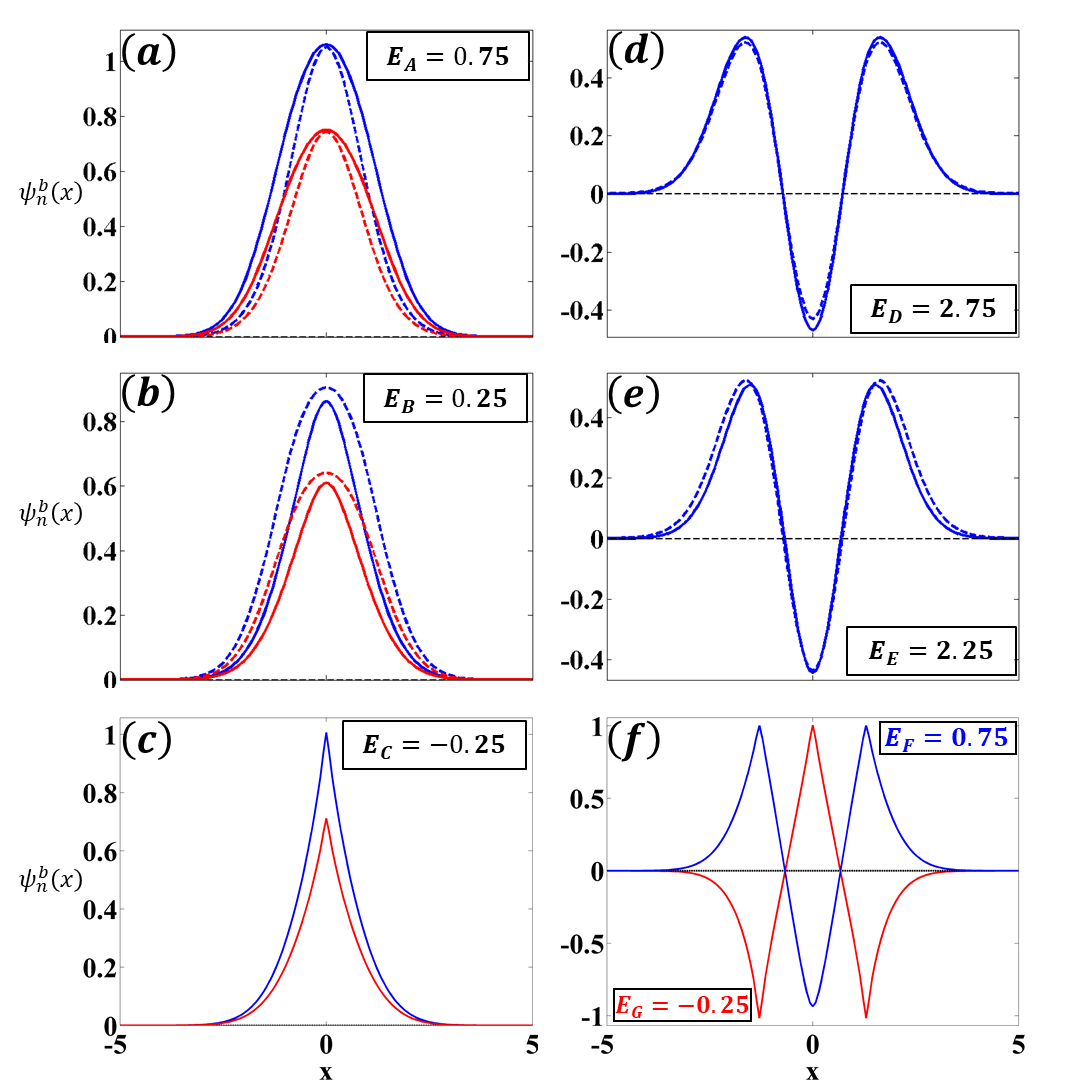}
\caption{The wavefunction for (a-c) the ground state $\psi_0^b(x)$ and (d-f) the second order excited state $\psi_2^b(x)$  of the quantum harmonic oscillator with DDEM $b = \pm 1$ and $b = \pm 2$, shown in Blue- and Red-colored curves, respectively, i.e., corresponding to the markers ($E_A$, $E_B$, $E_C$) and ($E_D$, $E_E$, $E_F$ and $E_G$) labelled in Figs. 1 and 3, respectively. 
The analytical results based on the perturbation theory are depicted in the dashed-curves, for $\psi_0^b(x)$ and $\psi_2^b(x)$, accordingly.
Here, the selected eigen-energies $E$ are chosen to represent the typical profile of wavefunctions  in three different regions: (a, d) a smooth profile with $b > 0$; (b, f) a smooth profile with $b < 0$, and (c, f) a peakon-like solution with $b < 0$.}
\end{figure}

However, when $b$ is negative, there are two distinct regions in this $Q$-$E$ curve, illustrated in the Green- and Yellow-colored backgrounds in Fig. 1.
For the Green-colored region, the corresponding eigen-energy is smaller than $E_0 = 0.5$, but remains positive, i.e., $0 < E_0^b < E_0$.
The probability factor $Q(E_0^b)$ is also linearly proportional to the eigen-energy $E$, as predicted by our analytical formula in Eq.~(\ref{sm}).
But, now the slope of $Q$-$E$ curve changes its sign, as $b < 0$.
The resulting wavefunction $\psi_0^b(x)$, as shown in Fig. 2(b) for the marked eigen-energy $E_B = 0.25$, still has a smooth profile.
However, the corresponding width of wavefunction shrinks when $E \to 0$.
As a result, a singularity emerges at $E_0^s = 0$ for the ground state, in which no well-defined localized wavefunction can be supported. 
The singularity comes from the divergence of $Q(E)$  near $1+b \vert  \psi \vert^2 $=0.  
Moreover, as one can see, our theoretical formula also breaks down when $E$ approaches this singularity.

Unexpectedly, single-hump solutions can be  supported even when $E < E_0^s = 0$, as show in the  Yellow-colored region. 
As shown in Fig. 2(c) for the marked eigen-energy $E_C = -0.25$, instead of a  smooth profile  stemmed from the Gauss wavefunction, the resulting wavefunction of this family solutions has a discontinuity in  their first-order derivative, similar to the {\it peakon} solution in the form of $\exp(-\vert x\vert)$.
Such peakon-like solutions are also already found in the IDD setting for optical waves, even without the introduction of harmonic oscillators~\cite{nlindispersion, IDD}.
As our perturbation theory starts from the eigen-basis of Hermite-Gaussian functions, it is not applicable to this family of peak-like solutions.

\begin{figure}
\includegraphics[width=8.6cm]{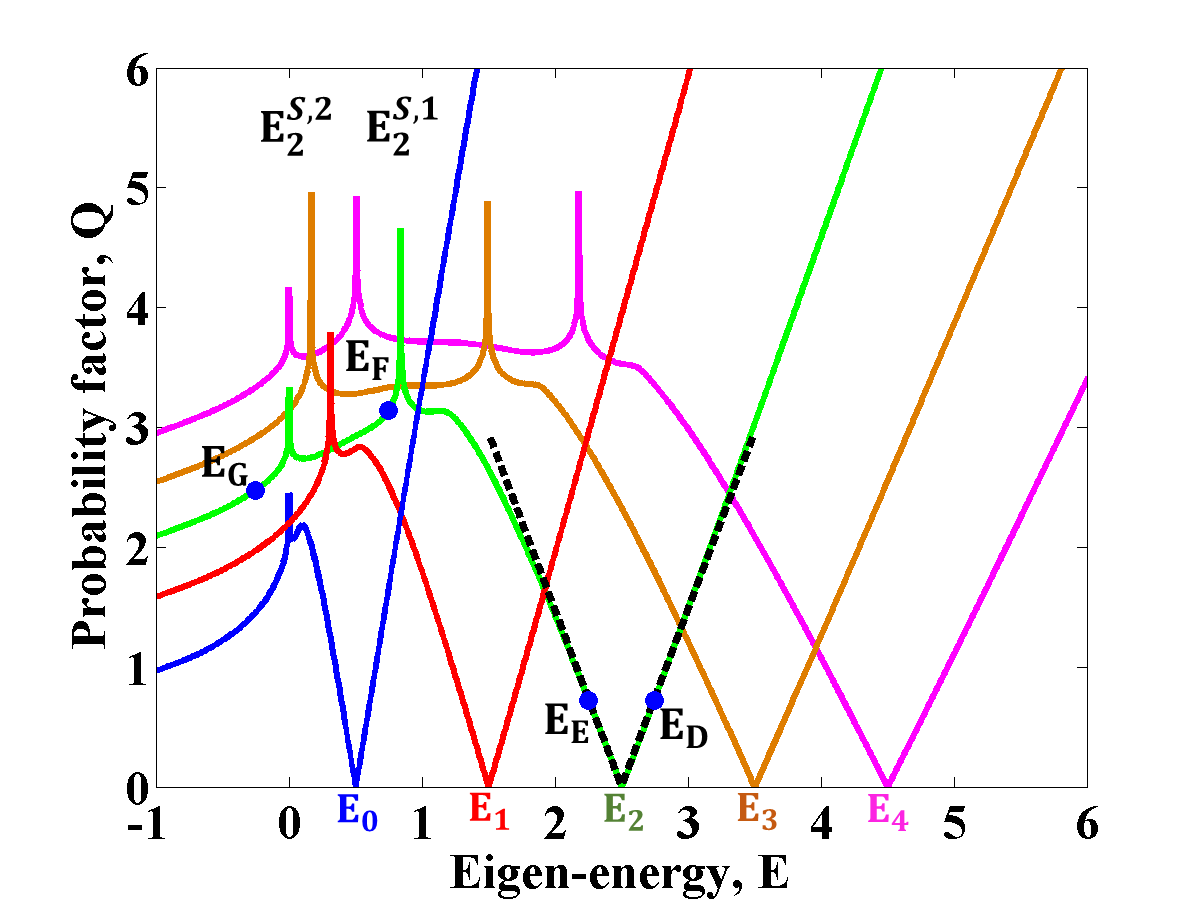}
\caption{The $Q$-$E$ curve, i.e., probability factor versus eigen-energy curve, for the lowest five states in a quantum harmonic oscillator with DDEM. Here, the DDEM parameter $b$ is set as $\pm 1$. The solution family with the same number of humps in the eigen-functions $\psi_n^b(x)$ are depicted in the same colors, with the labelled starting eigen-energies $E_n$, for  $n = 0, 1, 2, 3, 4$. 
Analytical solutions based on the perturbation theory given in Eq.~(\ref{sm2}) are also depicted in the Black dashed-curves for the second order excited states, which illustrate good agreement to the numerical solutions near $E_2 = 2.5$.}
\end{figure}

\subsection{The Excited States}
In addition to the ground state, the founded second order excited states, both numerically and analytically, are also depicted in Figs. 2(d-f) in solid- and dashed-curves, respectively.
Again, we also have three different regions in characterizing the wavefunction profiles.
Smooth profiles with the DDEM $b>0$ and $b < 0$ are shown in Figs. 2(d) and (e) for the marked eigen-energies $E_D = 2.75 > E_2 = 2.5$ and $E_E = 2.25 < E_2$ in Fig. 3, respectively.
As shown in Figs. 2(d) and (e), the two solutions, $\psi_2^b(x)$,  have three humps in their profiles and share the similar profile as the $2$nd order Hermite-Gaussian  function.
By comparing the solid- and dashed-curves, corresponding to our numerical results and analytical solutions, respectively, one can see nearly perfect agreement for the solutions around the eigen-energy $E_2$. 

Nevertheless, when $b<0$ and $E < E_2^{s,1} \approx 0.8398$,  a discontinuous profile emerges due to the singularity happened in the $Q$-$E$ curve.
Unlike the $Q$-$E$ curves for the ground state, there exist two singularities, denoted as $E_2^{s,1}$ and $E_2^{s,2} = 0$. 
When the eigen-energy is smaller than the first singular energy $E_2^{s,1}$ but larger than the second singular energy $E_2^{s, 2}$,  i.e., $E_2^{s,2} < E_2^b < E_2^{s,1}$, for example $E_F = 0.75$, the peakon-like solution illustrated in Blue-color in Fig. 2(f),  has a profile of  $\exp(-\vert x\vert)$ in two  of the humps in the sidebands.
It is noted that the profile in the central hump remains a smooth one.
Nevertheless, when the eigen-energy is smaller the value at the second singularity $E_2^{s, 2}$ , for example $E_G = -0.25$, the corresponding eigen-function has discontinuities in all the three humps, as the Red-colored curve depicted in Fig. 2(f).

In Fig. 3,  we plot all the founded eigen-energies, up to $n = 4$, by depicting the solution family with the same number of humps in the eigen-functions $\psi_n^b(x)$ in the same colors.
One can see clearly that, all the $Q$-$E$ curves start from the eigen-energies $E_n = n + \frac{1}{2}$ of a standard quantum harmonic oscillator, i.e., $b = 0$.
Around these energy values, $E_n$, our perturbation theory works perfectly, giving the linear dependence of  $Q(E)$ on the eigen-energy, along with the inversely proportional relation to the DDEM parameter $b$.
In particular, as depicted in the Black dashed-curves, our analytical solutions given in Eq.~(\ref{sm2}) also illustrate good agreement to the numerical solutions for the second order excited states.

However, when $b$ turns negative and the supported eigen-energy $E_n^b$ is away from the starting energy value $E_n$,  more singularities appear at certain value(s) of $E_n^s$.
The number of singularity depends on the critical points of the Hermite polynomial due to the divergence of $Q(E)$  near $1+b \vert  \psi \vert^2 $=0. 
Considering the symmetry of Hermite polynomial, i.e., $ H_n (x) =\pm H_n (-x)$, as one can see from Fig. 3,  the number of singularities for $Q(E_{2n}^b)$ and $Q(E_{2n+1}^b)$ is the same, i.e., equal to $n+1$.

Before Conclusion, we remark the stability of the founded eigen-solutions in our generalized quantum harmonic oscillator with a probability density-dependent effective mass (DDEM).
As confined by the external harmonic oscillator, all the found eigen-solutions are stable numerically.
The validity of our perturbation theory is limited to the eigen-energy around the known one $E_n = 2n+1$.
It is expected that our analytical formula breaks down when $E_n^b$ approaches the singular energy $E_n^s$.
As for the possible bifurcation maps, how to develop an analytical method to find the solutions for these peakon-like solutions, as well as around the singularities, remains a challenge, which goes beyond the scope of the current work but deserves further studies.

\section{Conclusion}
With the introduction of probability density-dependent effective mass (DDEM) for a quantum particle in harmonic oscillators, we propose a generalized Schr{\"o}dinger equation to embrace the nonlinear effective mass.
With the help of orthonormal property of Hermite-Gaussian  functions, we reduce this partial differential equation into an infinite discrete dynamical system and find the corresponding stationary solution by perturbation theory.
The monotonicity of perturbed solutions is also approved rigorously.
The resulting eigen-energy spectra is no long discretized, but continuous due to the introduction of a nonlinear effective mass.
With the comparison to numerical results obtained by direct simulations,  the validity of our analytical formula in the asymptotic limit, in terms of the probability factor as a function of the eigen-energy, $Q(E)$, can be easily verified, in particular for the solutions stemmed  from the expansion of Hermite-Gaussian  functions.
However, the nonlinear effective mass also introduces a new family of  peakon-like solutions with a discontinuity in  their first-order derivative, which definitely deserves further studies.

It has been well studied with the nonlinear Schr{\"o}dinger wave equation, or the Gross-Pitaevskii  equation in general, where the nonlinear terms come from Kerr-effect, or  the mean-field interaction.
With the eigen-energy and eigen-function illustrated in this work, our proposed generalized quantum harmonic oscillator opens an unexplored area for quantum particles with nonlinear effective masses. 
A number of promising applications and directions for further exploration may be identified when particles  accessing nonlinear correction to their effective mass.
Similar models related to our proposed generalized quantum harmonic oscillators, but in more complicated settings involve off-resonant self-induced transparency (SIT) solitons~\cite{SIT, SIT2} spatially-periodic refractivity doped with two-level systems (TLS)~\cite{TLA, TLA2}, 
electromagnetically-induced transparency (EIT) via via resonant dipole-dipole interactions~\cite{3level, 3level2}, and the continuum limit  of the Salerno model~\cite{Salerno}.

\subsection*{Acknowledgments}
This work is partially supported  by the Ministry of Science and Technology of Taiwan under Grant No.: 105- 2628-M-007-003-MY4, 107-2115-M-606-001, 108-2923-M-007-001-MY3, and 109-2112-M-007-019-MY3, as well as Office of Naval Research Global.

\appendix
\renewcommand\thesection{A}
\setcounter{equation}{0}
\section*{Appendix} \label{appA}
\begin{widetext}
\noindent Here, we give the details to prove the inequality shown in Eq.~(\ref{un}) and Eq.~(\ref{une}).

\noindent First of all, from  Eq. (\ref{re}), one can see that 
\bea  && W_{2n, 2n, 2n, 2n} -(4n+1) V_{2n, 2n, 2n, 2n} \\ \no
&=& \frac{\sqrt{2n(2n-1)}}{2}V_{2n-2,2n,2n,2n}-(2n+1/2) V_{2n,2n,2n,2n} 
+ \frac{\sqrt{(2n+1)(2n+2)}}{2}V_{2n+2,2n,2n,2n}, \\ \no
& < & n V_{2n-2,2n,2n,2n} -(2n+1/2) V_{2n,2n,2n,2n} + (n+1) V_{2n+2,2n,2n,2n} ,\\\no 
&=& n( V_{2n-2,2n,2n,2n}-V_{2n,2n,2n,2n}) + n ( V_{2n+2,2n,2n,2n}- V_{2n,2n,2n,2n}) 
+ (V_{2n+2,2n,2n,2n}-\frac{1}{2} V_{2n,2n,2n,2n}). \label{re2} \eea 
Then, with the formula 
\be \Gamma (h+\frac{1}{2})=  \left(\ba{c}  h+\frac{1}{2} \\ h   \ea \right) h! \sqrt{\pi}= \frac{(2h-1)!!}{2^h h!} h! \sqrt{\pi},  \label{fo} \ee
 one can have 
\be V_{2n, 2n, 2n, 2n}= \frac{1}{\sqrt{2 \pi}} \sum_{\nu=0}^{2n} \left(\ba{c} \nu-\frac{1}{2} \\ \nu   \ea \right) \left(\ba{c}  2n-\nu-\frac{1}{2} \\ 2n-\nu   \ea \right)^2,\label{q1} \ee
and
\[ V_{2(n-1), 2n, 2n, 2n}= \frac{1}{\sqrt{2 \pi}} \sum_{\nu=0}^{2(n-1)} \left(\ba{c} \nu-\frac{1}{2} \\ \nu   \ea \right) \left(\ba{c}  2n-\nu-\frac{1}{2} \\ 2n-\nu   \ea \right)^2\frac{(2n-\nu)(2n-\nu-1)}{(2n-\nu-\frac{1}{2})^2}.  \]
As the inequality $ \frac{(2n-\nu)(2n-\nu-1)}{(2n-\nu-\frac{1}{2})^2} <1 $ is hold, we can know that 
\bea  V_{2(n-1), 2n, 2n, 2n}   &<&  \frac{1}{\sqrt{2 \pi}} \sum_{\nu=0}^{2(n-1)} \left(\ba{c} \nu-\frac{1}{2} \\ \nu   \ea \right) \left(\ba{c}  2n-\nu-\frac{1}{2} \\ 2n-\nu   \ea \right)^2,  \\ \no
&=&V_{2n, 2n, 2n, 2n}  -\frac{1}{\sqrt{2 \pi}} \sum_{\nu=2n-1}^{2n} \left(\ba{c} \nu-\frac{1}{2} \\ \nu   \ea \right) \left(\ba{c}  2n-\nu-\frac{1}{2} \\ 2n-\nu   \ea \right)^2, \label{q2} \eea
as well as
\bea && V_{2(n+1), 2n, 2n, 2n} = \frac{1}{\sqrt{2 \pi}} \sum_{\nu=1}^{2n} \left(\ba{c} \nu-\frac{1}{2} \\ \nu   \ea \right) \left(\ba{c}  2n-\nu-\frac{1}{2} \\ 2n-\nu   \ea \right)^2\frac{2}{2\nu-1}\frac{(2n-\nu+\frac{1}{2})^2}{(2n-\nu+1)(2n-\nu+2)},\\ \no
&&\qquad <  \frac{1}{\sqrt{2 \pi}} [ \sum_{\nu=1}^{2} \left(\ba{c} \nu-\frac{1}{2} \\ \nu   \ea \right) \left(\ba{c}  2n-\nu-\frac{1}{2} \\ 2n-\nu   \ea \right)^2\frac{2}{2\nu-1}\frac{(2n-\nu+\frac{1}{2})^2}{(2n-\nu+1)(2n-\nu+2)} +\frac{1}{2} \sum_{\nu=3}^{2n} \left(\ba{c} \nu-\frac{1}{2} \\ \nu   \ea \right) \left(\ba{c}  2n-\nu-\frac{1}{2} \\ 2n-\nu   \ea \right)^2] \label{q4}. \eea
Moreover, as the inequality $  \frac{(2n-\nu+\frac{1}{2})^2}{(2n-\nu+1)(2n-\nu+2)} <1$ is also hold, we can have  
\bea && \sum_{\nu=1}^{2} \left(\ba{c} \nu-\frac{1}{2} \\ \nu   \ea \right) \left(\ba{c}  2n-\nu-\frac{1}{2} \\ 2n-\nu   \ea \right)^2\frac{2}{2\nu-1}\frac{(2n-\nu+\frac{1}{2})^2}{(2n-\nu+1)(2n-\nu+2)} \no \\
&& =\pi^{3/2} [\frac{(4n-5)!!}{2^{2n} (2n)! }]^2[ \frac{2n(4n-1)^2(4n-3)^2}{2n+1} + \frac{4n(2n-1)^2 (4n-3)^2 }{2n-1}],  \no \\
&& < \pi^{3/2} [\frac{(4n-5)!!}{2^{2n} (2n)! }]^2 [ (4n-1)^2(4n-3)^2 + 3 (2n-1)^2 (4n-3)^2], \no \\
&&= \pi^{3/2} [\frac{(4n-5)!!}{2^{2n} (2n)! }]^2 ( 448n^4-992n^3+796n^2-276n+36 ).
\label{q5} \eea
Then, with the fact that  
\bea 
 \sum_{\nu=0}^{2} \left(\ba{c} \nu-\frac{1}{2} \\ \nu   \ea \right) \left(\ba{c}  2n-\nu-\frac{1}{2} \\ 2n-\nu   \ea \right)^2 = \pi^{3/2} [\frac{(4n-5)!!}{2^{2n} (2n)! }]^2[  576 n^4-896 n^3+472 n^2-96 n+9], \label{q6} \eea
from Eqs.~(\ref{q5}) and (\ref{q6}), one can reach at the following inequality: 
\bea  &&  \sum_{\nu=1}^{2} \left(\ba{c} \nu-\frac{1}{2} \\ \nu   \ea \right) \left(\ba{c}  2n-\nu-\frac{1}{2} \\ 2n-\nu   \ea \right)^2\frac{2}{2\nu-1}\frac{(2n-\nu+\frac{1}{2})^2}{(2n-\nu+1)(2n-\nu+2)}  < \sum_{\nu=0}^{2} \left(\ba{c} \nu-\frac{1}{2} \\ \nu   \ea \right) \left(\ba{c}  2n-\nu-\frac{1}{2} \\ 2n-\nu   \ea \right)^2, \label{q7} \eea
when $n \geq 1$. Consequently, combining Eqs.~(\ref{q4}) and (\ref{q7}), we have 
\bea &&V_{2(n+1), 2n, 2n, 2n} < \frac{1}{2} V_{2n, 2n, 2n, 2n}+ \frac{1}{2\sqrt{2\pi}}\sum_{\nu=0}^{2} \left(\ba{c} \nu-\frac{1}{2} \\ \nu   \ea \right) \left(\ba{c}  2n-\nu-\frac{1}{2} \\ 2n-\nu   \ea \right)^2. 
\label{q8} \eea 
With the results obtained in  Eqs.~(\ref{re2}), (\ref{q2}) and (\ref{q8}), the inequality shown in Eq.~(\ref{une}) can be reached
\bea && W_{2n, 2n, 2n, 2n} -(4n+1) V_{2n, 2n, 2n, 2n}  \label{q9} \no\\
&&<  \frac{-n}{\sqrt{2\pi}}\sum_{\nu=2n-1}^{2n} \left(\ba{c} \nu-\frac{1}{2} \\ \nu   \ea \right) \left(\ba{c}  2n-\nu-\frac{1}{2} \\ 2n-\nu   \ea \right)^2 -\frac{n}{2\sqrt{2\pi}} \sum_{\nu=3}^{2n} \left(\ba{c} \nu-\frac{1}{2} \\ \nu   \ea \right) \left(\ba{c}  2n-\nu-\frac{1}{2} \\ 2n-\nu   \ea \right)^2  \\ \no
&&\quad+ \frac{n}{2\sqrt{2\pi}}\sum_{\nu=0}^{2} \left(\ba{c} \nu-\frac{1}{2} \\ \nu   \ea \right) \left(\ba{c}  2n-\nu-\frac{1}{2} \\ 2n-\nu   \ea \right)^2. \eea
It is noted that the last two terms shown in Eq. (\ref{q9}) is negative when $n \geq 2$. 

\noindent This completes the proof. 
\end{widetext}

\end{document}